\documentclass[12pt]{article}

\begin{document}

\input amssym.tex

\title{Discrete $O(1,4)$ transformations and new scalar quantum
modes on the de Sitter spacetime}

\author{Ion I. Cot\u aescu \thanks{E-mail:~~cota@physics.uvt.ro}\\
{\it West University of Timi\c soara,}\\{\it V. Parvan Ave. 4,
RO-300223 Timi\c soara, Romania}}

\maketitle

\begin{abstract}
It is shown that the isometry group of the de Sitter spacetime
includes two different three-dimensional Abelian subgroups which
transform between themselves through a discrete isometry
corresponding to the time reversal in the five-dimensional Minkowski
spacetime embedding the de Sitter one. The eigenfunctions of the
generators of these Abelian subgroups form two different sets of
quantum modes correlated by the mentioned isometry.

Pacs: 04.20.Cv, 04.62.+v, 11.30.-j
\end{abstract}

\vspace*{12mm} Keywords: de Sitter; isometries; discrete isometries;
scalar quantum modes.

\newpage

The quantum theory of fields on the de Sitter spacetime has the
advantage of the $SO(1,4)$ symmetry \cite{SW} which provides one
with the conserved $so(1,4)$ generators  commuting with the
operators of the field equations. These differential operators
represent the principal quantum observables which enable one to
define quantum modes as common eigenfunctions of several sets of
commuting operators including that of the field equation. However,
the physical meaning of these generators is not fully understood
even though in the flat limit one recovers the Poincar\' e symmetry.
An open problem is related to the presence in $SO(1,4)$ of two
Abelian subgroups $T(3)$ among them only one can be interpreted as
the subgroup of space translations in the (co)moving charts
\cite{BD}. The role of the second $T(3)$ subgroup remains obscure as
long as its generators are less studied so far. For this reason we
would like to analyze here the relation between these two $T(3)$
subgroups investigating the properties of the mode functions defined
as eigenfunctions of their generators. Our principal result points
out that these subgroups transform between themselves through a {\em
discrete} $O(1,4)$ de Sitter isometry corresponding to the time
reversal in the five-dimensional Minkowski spacetime embedding the
de Sitter one.

Let us start with the de Sitter spacetime $(M,g)$ defined as the
hyperboloid of radius $1/\omega$ \footnote{We denote by $\omega $
the Hubble de Sitter constant since  $H$ is reserved for the
Hamiltonian operator} in the five-dimensional flat spacetime
$(M^5,\eta^5)$ of coordinates $z^A$  (labeled by the indices
$A,\,B,...= 0,1,2,3,4$) and metric $\eta^5={\rm
diag}(1,-1,-1,-1,-1)$ \cite{SW}. A local chart of coordinates
$x^{\mu}$ ($\mu,\nu,...=0,1,2,3$) can be introduced on $(M,g)$
giving the set of functions $z^A(x)$ which solve the hyperboloid
equation, $\eta^5_{AB}z^A(x) z^B(x)=-\omega^{-2}$. In this manner
$(M,g)$ is defined  as a homogeneous space of the pseudo-orthogonal
group $O(1,4)$. The proper $O(1,4)$ transformations form the
subgroup $SO(1,4)$ while the improper ones have to be written using
discrete transformations on $M^5$.

The group $SO(1,4)$ is in the same time the gauge group of the
metric $\eta^{5}$ and the isometry group, $I(M)$, of the de Sitter
spacetime. Its universal covering group, $S(M)={\rm
Spin}(\eta^5)=Sp(2,2)$, is the group of external symmetry as defined
in Ref. \cite{ES}. This has the Lie algebra $s(M)=sp(2,2)\sim
so(1,4)$ for which we use the covariant real parameters
$\xi^{AB}=-\xi^{BA}$. In this parametrization,  any $\Lambda(\xi)\in
SO(1,4)$ produces the linear transformation $z^A(x)\to
z^{A}(x')=\Lambda(\xi)^{A\,\cdot}_{\cdot\,B}z^B(x)$  giving rise to
the isometry $x \to x'=\phi_{\xi}(x)=x+k_{AB}(x)\xi^{AB}+...$ which
can be expanded in terms of Killing vectors (of components
$k_{AB}^{\mu}$) associated to the parameters $\xi^{AB}$. These
Killing vectors allow one to write down the basis-generators of the
covariant representations according to which the matter fields
transform under isometries \cite{CML,ES,CCC}. In the simplest case
of the scalar fields $\psi$, which transform according to the
natural representation $\psi \to \psi'=\psi\circ \phi_{\xi}^{-1}$,
these generators are the orbital operators
$L_{AB}=-ik_{AB}^{\mu}\partial_{\mu}$ \cite{PPP,Cs}.

We assume now that $(M,g)$ is equipped with the local chart
$\{t,\vec{x}\}$ of {\em conformal} time $t$ and Cartesian space
coordinates defined by the functions
\begin{eqnarray}
z^0(x)&=&-\frac{1}{2\omega^2 t}
\left[1-\omega^2\left(t^2-{\vec{x}\,}^2\right)\right]\,,\label{z0}\\
z^4(x)&=&-\frac{1}{2\omega^2
t}\left[1+\omega^2\left(t^2-{\vec{x}\,}^2\right)\right]\,,\label{z4}\\
z^i(x)&=&-\frac{x^i}{\omega t}\,, \quad i,j,...=1,2,3\label{zi}\,,
\end{eqnarray}
giving rise to the conformal-flat line element
\begin{equation}\label{FRW}
ds^2=\eta^5_{AB}dz^A dz^B=\frac{1}{\omega^2
t^2}\,(dt^2-d\vec{x}\cdot d\vec{x})\,.
\end{equation}
This chart  covers the expanding part of $M$ for $t \in (-\infty,0)$
and $\vec{x}\in {\Bbb R}^3$ while the collapsing part is covered by
a similar chart with $t >0$ \cite{BD}.

In these charts it is convenient to use the basis $\{
H,P_i,Q_i,L_i\}$ of the natural representation of the $so(1,4)$
algebra which is formed by the energy operator $H \equiv  \omega
L_{(04)}=-i\omega(t\,\partial_t+ {x}^i {\partial}_i)$, the momentum
operator $\vec{P}$ and its dual, $\vec{Q}$, whose components are
defined as \cite{CCC}
\begin{eqnarray}
P_i\equiv \omega(L_{(i4)}-L_{(i0)})&=&i\partial_i\,,\label{Pi}\\
Q_i \equiv \omega(L_{(i4)}+L_{(i0)})&=&2i x^i \omega^2(t\partial_t+
{x}^j {\partial}_j)+
i\omega^2(t^2-{\vec{x}\,}^2)\partial_i\,,\label{Qi}
\end{eqnarray}
and the angular momentum $L_i \equiv  \frac{1}{2}\,\varepsilon_{ijk}
L_{(jk)}=-i\varepsilon_{ijk}x^j\partial_k$. These generators satisfy
the commutation relations \cite{CCC}
\begin{eqnarray}
&&\left[ H, P_i \right]=i\omega P_i\,,\quad \hspace*{8 mm} \left[ H,
Q_i
\right]=-i\omega Q_i\,,\label{HPQ}\\
&&\left[L_i , P_j \right]=i\varepsilon_{ijk} P_k\,,\hspace*{7mm}
\left[ L_i,
Q_j\right]=i\varepsilon_{ijk} Q_k\,,\label{PJP}\\
&&[P_i,P_j]=0\,, \hspace*{17mm} \left[Q_i, Q_j \right]=0\,,\\
&&\left[Q_i, P_j \right]=2 i \omega \delta_{ij} H + 2 i \omega^2
\varepsilon_{ijk} L_k \label{QiPi}\,.
\end{eqnarray}
The first Casimir operator of this algebra coincides to the
Klein-Gordon operator, ${\cal C}_1=H^2+3 i \omega
H-{\vec{Q}}\cdot{\vec{P}} -\omega^2{\vec{L}\,}^2={\cal E}_{KG}$,
while the second one, ${\cal C}_2=0$, vanishes since the there is no
spin \cite{CCC}. For this reason the usual scalar quantum modes,
called here $P$-modes, are determined as common eigenfunctions of
the set of commuting operators $\{ {\cal E}_{KG},P_i\}$ \cite{Cs}.

On the other hand, it is remarkable  that the operators $Q_i$ form
the subalgebra of an Abelian subgroup $T(3)_Q$ isomorphic to the
subgroup  $T(3)_P$ of the {\em physical} translations in the chart
$\{t,\vec{x}\}$ generated by $P_i$. This suggests us that new
quantum modes, called $Q$-modes, could be defined as common
eigenfunctions of the set $\{{\cal E}_{KG},Q_i\}$. These may be
written down by using a coordinate transformation able to change the
Abelian generators among themselves ($P_i \leftrightarrow Q_i$)
without to affect the other generators. This transformation can not
be an $SO(1,4)$ isometry since the $so(1,4)$ generators $L_{AB}$
transform under isometries as skew-symmetric tensors. This means
that we are left only with the discrete transformations of the group
$O(1,4)$ which need our attention in what follows.

The simplest discrete transformations on $M^5$, denoted by
${\Pi}_{[A]}$, are those which change the sign of a single
coordinate, $z^A\to -z^A$. These transformations give rise to the
discrete isometries $x\to x'=\pi_{[A]}(x)$ defined as $\Pi_{[A]}
z(x)=z[\pi_{[A]}(x)]$. Obviously, these isometries satisfy
$\pi_{[A]}\circ\pi_{[A]}=id$ where $id$ denote the identity
function. A rapid inspection indicates that the interesting
non-trivial isometries are produced by $\Pi_{[0]}$ as
\begin{equation}\label{Pi0}
t'=\pi_{[0]}^0(x)=\frac{t}{\omega^2 (t^2-{\vec{x}\,}^2)}\,, \quad
x^{i\,\prime}=\pi_{[0]}^i(x)=\frac{x^i}{\omega^2
(t^2-{\vec{x}\,}^2)}\,,
\end{equation}
and by $\Pi_{[4]}$ which gives
\begin{equation}\label{Pi4}
t'=\pi_{[4]}^0(x)=-\frac{t}{\omega^2 (t^2-{\vec{x}\,}^2)}\,, \quad
x^{i\,\prime}=\pi_{[4]}^i(x)=-\frac{x^i}{\omega^2
(t^2-{\vec{x}\,}^2)}\,.
\end{equation}
The other isometries, $\pi_{[i]}$, are simple mirror transformations
of the space coordinates $x^i$ such that the (space) parity reads
$\pi_{[\vec{x}]}=\pi_{[1]}\circ\pi_{[2]}\circ\pi_{[3]}$. Another
remarkable discrete isometry is $\pi_{[x]}=\pi_{[0]}\circ\pi_{[4]}$
which changes the signs of all the coordinates $x^{\mu}$. It is
worth noting that the physical measurements can be performed only
inside the light-cone where $|t|>|\vec{x}|$. This indicates that the
isometry (\ref{Pi0}) does not change the signs of the time and space
coordinates and, consequently, the charts $\{t,\vec{x}\}$ and
$\{t',{\vec{x}\,}'\}$ cover the same portion of $M$. On the
contrary, the isometry (\ref{Pi4}) changes the sign of the conformal
time moving the transformed chart to the opposite portion of this
manifold.

Now our problem can be solved observing that the transformation
$\Pi_{[0]}$ changes only the signs of the generators $L_{(0i)}$ and
$L_{(04)}$ so that, in the chart $\{t',{\vec{x}\,}'\}$ defined by
equations (\ref{Pi0}), we meet the new basis-generators
\begin{equation}\label{cucu}
P'_i=Q_i\,,\quad Q'_{i}=P_i\,,\quad H'=-H\,, \quad L'_i=L_i\,,
\end{equation}
but the same Casimir operator ${\cal E}'_{KG}={\cal E}_{KG}$ since
$3 i \omega H'-{\vec{Q}\,}'\cdot{\vec{P}\,}' = 3 i \omega
H-{\vec{Q}}\cdot{\vec{P}}$ as it results from equations (\ref{QiPi})
and (\ref{cucu}). Therefore, the $P$-mode functions $f_{\vec{p}}(x)$
of the chart $\{t,{\vec{x}}\}$, which satisfy the Klein-Gordon
equation ${\cal E}_{KG}f_{\vec{p}}(x)=m^2 f_{\vec{p}}(x)$ and the
eigenvalue problems $P_i f_{\vec{p}}(x)=p_i f_{\vec{p}}(x)$, become
$Q$-mode functions in the chart $\{t',{\vec{x}\,}'\}$ where these
obey ${\cal E}'_{KG}f_{\vec{p}}(x')=m^2 f_{\vec{p}}(x')$ and $Q'_i
f_{\vec{p}}(x')=p_i f_{\vec{p}}(x')$. We arrive thus to our
principal conclusion:\\

{\em Given the set of $P$-mode functions $f_{\vec{p}}$ in the chart
$\{t,{\vec{x}}\}$, the functions $f_{\vec{q}}\circ \pi_{[0]}$ form
the set of $Q$-mode functions of the same chart, depending on the
parameters $\vec{q}\in {\Bbb R}^3$.}\\

\noindent Obviously, these new mode functions are solutions of the
Klein-Gordon equation and satisfy $Q_i f_{\vec{q}}[\pi_{[0]}(x)]=q_i
f_{\vec{q}}[\pi_{[0]}(x)]$ where the operators $Q_i$ are given by
equation (\ref{Qi}).

Hence our problem is completely solved but this gives rise to other
questions to be addressed in further investigations. It remains to
study what happens with the similar modes of the fields with spin
and how the discrete isometries act in this case. Moreover, the
physical meaning of the $Q$-modes must be elucidated by working out
significant examples.

\subsection*{Acknowledgements}

This work is partially supported by the ICTP-SEENET-MTP grant PRJ-09
in frame of the SEENET-MTP Network G. S. Dj..

\end{document}